\documentclass[aps,amsmath,amssymb,showpacs,pre]{revtex4-1}
\newcommand{\be}{\begin{equation}}
\newcommand{\ee}{\end{equation}}
\newcommand{\bea}{\begin{eqnarray}}
\newcommand{\eea}{\end{eqnarray}}
\usepackage{epsfig}
\usepackage{bm}
\usepackage{latexsym}
\usepackage{graphicx}
\usepackage{comment}

\usepackage{graphicx,color}
 \graphicspath{{./}{./figures/}}
\usepackage{enumerate}
\usepackage{mathbbol}
\usepackage{amsfonts}
\usepackage{natbib}

\usepackage{bm}
\usepackage{hyperref}

\usepackage{color}

\begin{document}

\title{Path integrals for higher derivative actions }
\author{David S. Dean$^{1,2}$, Bing Miao$^{3}$, Rudi Podgornik$^{2,4,5}$}
\affiliation{(1) Univ. Bordeaux, CNRS, LOMA, UMR 5798, F-33400 Talence, France}
\affiliation{(2) School of Physical Sciences and Kavli Institute for Theoretical Sciences, University of Chinese Academy of Sciences, Beijing 100049, China}
\affiliation{(3) Center of Materials Science and Optoelectronics Engineering, \\ College of Materials Science and Opto-Electronic Technology, University of Chinese Academy of Sciences, Beijing 100049, China}
\affiliation{(4)CAS Key Laboratory of Soft Matter Physics, Institute of Physics,
Chinese Academy of Sciences (CAS), Beijing 100190, China}
\affiliation{(5) Department of Physics, Faculty of Mathematics and Physics, University of Ljubljana,  1000 Ljubljana, Slovenia}
\bibliographystyle{apsrev}
\begin{abstract}{We consider Euclidean path integrals with higher derivative actions, including those that depend quadratically on acceleration, velocity and position. Such path integrals arise naturally in the study of stiff polymers, membranes with bending rigidity as well as a number of models for electrolytes. The approach used is based on the relation between quadratic path integrals and Gaussian fields and we also show how it can be extended to the evaluation of even higher order path integrals. }
\end{abstract}
\maketitle
\section{Introduction}
In this paper  we revisit the problem of computing  path integrals with actions containing higher derivatives, such as
\begin{equation}
K(x,v,x',v';t) = \int_{x(0)=x,\dot x(0)=v} d[x] \delta(\dot x(t)-v')\delta(x(t)-x')
\exp\left(-\frac{1}{2}\int_0^t  ds \ [ \frac{d^2 x(s)}{ds^2}]^2 + m  [ \frac{dx(s)}{ds}]^2 + k x^2(s)\right).
\end{equation}
The  action in the path integral is {\rm restricted} to the interval $[0,t]$, as denoted by the integration limits over the variable $s$, implying that the fluctuating field does not exist outside of the interval $[0,t]$. The above path integral was first evaluated by Kleinert \cite{klein86} and the result is also given in one of the standard text books on path integration \cite{handbook}. However some special cases of this result were established before \cite{pap}.

Path integrals of the above type have a number of applications. They are of course relevant to stiff or semi-flexible polymers \cite{pap,freed,doi,dob2001,smith,uchida} whose overall length can vary. They also arise in the study of membranes with bending rigidity \cite{dean2007} as well as liquid crystal theory \cite{had2018}. In the context of quantum mechanics they also arise naturally when weak relativistic corrections to standard quantum mechanics are taken into account \cite{simon1990}. Recent developments in the theory of ionic liquids, where finite ion sizes are important have proposed mean field theories which introduce higher order derivatives than the usual second order Poisson-Boltzmann theory \cite{sant2006,bazant2011,blossey2017} and the analysis of the one loop correction to such theories will require higher order path integral formulations. In addition, field theories with higher derivative, and even non-local, actions also arise from the dynamics of lower derivative theories when analysing the evolution  of Casimir forces toward their equilibrium values from non-equilibrium initial states \cite{ddaj1,ddaj2}.

The difference with respect to usual path integrals is the presence of the term $\frac{1}{2}\int_0^t  ds [ \frac{d^2 x(s)}{ds^2}]^2$ in the measure of the path integral (whose coefficient without loss of generality we set to be $1$). In polymer language this term corresponds to a bending energy while the standard Gaussian elastic chain model only possesses a monomer bond elastic energy of the form $\frac{1}{2}\int_0^t  ds\  m \  [ \frac{dx(s)}{ds}]^2$. Specifically, this last term in the Gaussian elastic model corresponds to the Wiener measure on the increments of the polymer position. These increments are independent and uncorrelated, ensuring that the polymer path $x(s)$ is continuous while the velocity remains discontinuous. In the absence of the confining quadratic potential $kx^2/2$, the Gaussian elastic model trajectory is equivalent to the Brownian motion. However for the model with a bending energy the measure on the velocity is that of Brownian motion and the velocity itself is therefore continuous. As the velocity $dx(s)/ds$ is continuous, one therefore needs to specify its value $v$ at the starting point $s=0$ and $v'$ at the end point $s=t$.

Strictly speaking therefore the measure in the above path integral is a Wiener measure on the velocity $v= dx/ds$ and the propagator should, more appropriately in a mathematical sense, be written as
\begin{equation}
K(x,v,x',v';t) = \int_{x(0)=x,\dot x(0)=v} d[v] \delta(\dot x(t)-v')\delta(x(t)-x')
\exp\left(-\frac{1}{2}\int_0^t  ds [ \frac{d v(s)}{ds}]^2 + (\omega_1^2+\omega_2^2)  v(s)^2 + \omega_1^2 \omega_2^2 x^2(s)\right).
\end{equation}
where $x(s) = x(0) +\int_0^s du\  v(u)$.  Here, for convenience, we have introduced the fundamental frequencies $\omega_1$ and $\omega_2$ defined by
$\omega_1^2 + \omega_2^2 = m$ and $\omega_1^2\omega_2^2= k$. In probabilistic terms one can thus write
\begin{equation}
K(x,v,x',v';t) = {\mathbb E}^{v,x}\left(\delta(x(t)-x')\delta(v(t)-v')\exp[-\frac{1}{2}\int_0^t
(\omega_1^2+\omega^2_2) v^2(s) + \omega_1^2\omega_2^2 x^2(s)]
\right),
\end{equation}
where ${\mathbb E}^{v,x}$ denotes the average over the the process $v(s),\ x(s)$ which obeys the stochastic differential equation
\begin{eqnarray}
dx(s) &=& v(s) ds \\
dv(s) &=& dB_s,
\end{eqnarray}
where $dB_s$ is a standard Brownian motion such that $dB_s^2 = ds$. Furthermore, the process is taken with initial conditions $x(0)=x$ and $v(0)=v$. It is well known, and indeed to some extent is the whole point of path integrals, that the standard path integrals 
are solutions to the Euclidean Schr\"odinger equation as stated by the Feynman-Kac formula.

The purpose of the theory we present here is, that it can be straightforwardly extended  to path integrals with actions containing higher derivative terms of the form
\begin{equation}
K({\bf X},{\bf X}';t) = \int_{{\bf X(0)={\bf X}}} d[x] \delta({\bf X}(t)-{\bf X}')
\exp\left(-\frac{1}{2}\int_0^t  ds \sum_{k=0}^n a_k \left[\frac{d^kx(s)}{ds^k}\right]^2\right),
\end{equation}
where, without any loss of generality, we systematically set $a_n=1$ and where we use the compact notation
\begin{equation}
{\bf X}(s) = \begin{pmatrix} & x(s)\\& \dot x(s) \\& \ddot x(s) \\& \cdot \\&\cdot \\& x^{(n-1)}(s)\end{pmatrix}.
\end{equation}
Although the algebraic complexity, and the resulting formulas increases considerably on increasing the order of the path integral, the method described here can be systematically applied and always leads to explicit results.

The paper is organised as follows, in section (\ref{fc1}) we show how the path integral can be written as a solution to a partial differential (Feynman-Kac) equation, the equivalent of  a Schr\"odinger equation for standard path integrals. For completeness the Feynman-Kac formula is derived in a heuristic manner. In section (\ref{fc2}) we show that the time  dependent solution of this equation can be seen to be of a Gaussian form, but the time dependent coefficients obey nonlinear equations which are very difficult to solve. However, we show that the late time, {\sl ground state dominance},  solution to the Schr\"odinger equation can be identified as the solution of an algebraic Riccati equation \cite{riccati}.  In section (\ref{gaus}) we examine another path integral which corresponds to a free Gaussian theory which is invariant by translation in time as the field is present at all times. We call this the unconfined path integral, due to the presence of the field outside the interval $[0,t]$. When the fields are fixed at the times $0$ and $t$, the path integral corresponds to the joint probability density function of the field and its derivatives up to order $n-1$ and can be computed using a standard results on the distribution of Gaussian random variables.  In section (\ref{trans}) we then show how the confined propagator and the unconfined propagator can be related. This uses the results of section (\ref{fc2}) on the later time behavior of the confined propagator. In section (\ref{klein}) we show how this formalism leads to the path integral for a second order theory, recovering the result of Kleinert \cite{klein86} and paving the way to other higher order actions. The earlier developments however did not assume that the path integral is Gaussian in nature and that formally it remains Gaussian even for general $n^{th}$ order path integrals. In section \ref{higher} we show how the proposed methodology, can extend  Kleinert's results  to third order path integrals. Potential applications of such results are in polymer physics. If one is interested in higher order elastic effects for instance, the quadratic second order  derivative term (or local curvature term corresponding to the bending energy) in the action can be seen to induce a nonzero finite persistence length in the tangent vector of the polymer. A quadratic term in the third order derivative, a so called torsional form in \cite{freed}, would then induce a finite persistence length for the local curvature. In addition, as mentioned above, higher order derivative terms arise naturally when considering the dynamics of Gaussian fields in the context of the Casimir effect and the order depends on the nature of the dynamics (for example conserved - model A- or nonconserved - model B - dynamics).
\section{Feynman-Kac formula}\label{fc1}
In probabilistic terms the correspondence between the Schr\"odinger equation and the path integral formulations is known as the {\sl Feynman-Kac formula} , that indeed holds for a wide class of stochastic differential equations. Here we give a quick informal derivation based on some simple notions of stochastic calculus, making the derivation analogous to the corresponding derivation of the Fokker-Planck equation for the process. We start by  considering
\begin{equation}
A_f({\bf X}, t) = {\mathbb E}^{{\bf X}} \left( f({\bf X_t}) \exp(-\int_0^t ds\ V({\bf X}_s))\right)
\end{equation}
where ${\bf X} = {\bf X}(0)$ denotes the initial position of a stochastic process ${\bf X}$. Looking at how $A_f({\bf X}, t)$ changes between $t$ and $t+dt$ we find
\begin{eqnarray}
dA_f({\bf X}, t)&=& A_f({\bf X}, t+dt )- A_f({\bf X}, t)
= {\mathbb E}^{{\bf X}}\left( f({\bf X_t}+ d{\bf X}_t) \exp(-\int_0^{t+dt} ds\ V({\bf X}_s) - f({\bf X_t}) \exp(-\int_0^t ds\ V({\bf X}_s) )\right) \nonumber \\
&=& {\mathbb E}^{{\bf X}}\left( \left[ \nabla_i f({\bf X}_t) dX_{it}
+\frac{1}{2} \nabla_i\nabla_j f({\bf X}_t) dX_{it} dX_{jt} - dt V({\bf X}_t)f({\bf X}_t)\right]
\exp(-\int_0^t ds\ V({\bf X}_s))\right),
\end{eqnarray}
where we have used the Ito prescription of the stochastic calculus. The partial differential operator, known as the generator,  $G$, of the stochastic process, is then defined via the average over the increments $dX_{it}$, leading to

\begin{equation}
{\mathbb E}^{{\bf X}}\left[ \nabla_i f({\bf X}_t) dX_{it}
+\frac{1}{2} \nabla_i\nabla_j f({\bf X}_t) dX_{it} dX_{jt}\right]
= {\mathbb E}^{{\bf X}}\left[G f({\bf X}_t) dt\right].
\end{equation}

We thus find that
\begin{equation}
\frac{\partial A_f({\bf X}, t)}{\partial t} = {\mathbb E}^{{\bf X}}\left( \left[ Gf({\bf X}_t) -  V({\bf X}_t)f({\bf X}_t)\right]
\exp(-\int_0^t ds\ V({\bf X}_s))\right)
\label{fk1}
\end{equation}
Next, we consider the propagator
\begin{equation}
K({\bf X},{\bf X}'; t) = {\mathbb E}^{{\bf X}} \left( \delta({\bf X_t}-{\bf X'}) \exp(-\int_0^t ds\ V({\bf X}_s)\right).
\end{equation}
so that
\begin{equation}
A_f({\bf X}, t)= \int d{\bf X}' K({\bf X},{\bf X}'; t)f({\bf X}'),
\end{equation}
and Eq. (\ref{fk1}) can be written as
\begin{equation}
\int d{\bf X}' \frac{\partial K({\bf X},{\bf X}'; t)}{\partial t}f({\bf X}')
= \int d{\bf X}' K({\bf X},{\bf X}'; t) [Gf({\bf X'})-V({\bf X}')f({\bf X'})],
\end{equation}
where the generator $G$ operates on the variables ${\bf X}'$.
Now introducing the adjoint of the generator $G$ denoted by $G^\dagger$ we find
\begin{equation}
\int d{\bf X}' \frac{\partial K({\bf X},{\bf X}'; t)}{\partial t}f({\bf X}')
= \int d{\bf X}' f({\bf X}') [G^\dagger K({\bf X},{\bf X}'; t) -V({\bf X}')K({\bf X},{\bf X}'; t)].
\end{equation}
As the above is true for all suitably well behaved functions, $f$, we therefore derive the {\sl forward Feynman-Kac equation}
\begin{equation}
\frac{\partial K({\bf X},{\bf X}'; t)}{\partial t}= G^\dagger K({\bf X},{\bf X}'; t) -V({\bf X}')K({\bf X},{\bf X}'; t),
\end{equation}
with the initial condition
\begin{equation}
K({\bf X},{\bf X}';0) = \delta({\bf X}-{\bf X}').
\end{equation}
The Feynman-Kac equation in terms of the initial position ${\bf X}$ can be
derived by considering what happens in the first time step $dt$ after $t=0$, where the process moves by ${\bf dX}_0$. This means that, when $V$ only depends on time through ${\bf X}_t$ we can write
\begin{equation}
K({\bf X},{\bf X}'; t) = {\mathbb E}^{d{\bf X}_0}\left( \delta({\bf X}_{t-dt} + d{\bf X}_0-{\bf X'})\exp\left(-\int_{0}^{t-dt} V({\bf X}_s) dt\right)(1- V({\bf X})dt)\right),
\end{equation}
that is to say after $dt$ the process starts at the new initial position ${\bf X}+d{\bf X}_0$ and runs for a total time $t-dt$. The factor of $(1- V({\bf X})dt)$ comes from the initial step.
Note that if the potential $V$ is set to zero we simply recover the standard forward Fokker-Planck equation for the probability density function $K({\bf X},{\bf X'};t) = P({\bf X},{\bf X'};t)$ for probability density function at ${\bf X}'$, at time $t$,  of the process started at ${\bf X}$. This thus gives
\begin{equation}
K({\bf X},{\bf X}'; t)  = {\mathbb E}^{d{\bf X_0}}\left(
K({\bf X},{\bf X'}+ d{\bf X}_0; t-dt)(1-V({\bf X}) dt)\right).
\end{equation}
Expanding this to order $dt$ using the definition of the generator $G$ we find
\begin{equation}
K({\bf X},{\bf X}'; t) = K({\bf X},{\bf X}'; t)  (1- V({\bf X}) dt) - \frac{\partial K({\bf X},{\bf X}'; t)}{\partial t} dt + GK({\bf X},{\bf X}'; t)dt,
\end{equation}
where here $G$ now acts on the initial coordinate ${\bf X}$. This thus yields the backward Feynman-Kac equation
\begin{equation}
\frac{\partial K({\bf X},{\bf X}'; t)}{\partial t}= GK({\bf X},{\bf X}'; t)- V({\bf X})K({\bf X},{\bf X}'; t).
\end{equation}

\section{Large time properties of the Feynman-Kac propagator}\label{fc2}
A slightly more general case than what we consider here is the equation
\begin{equation}
\frac{\partial K({\bf X},{\bf X}'; t)}{\partial t}= \frac{1}{2}D_{ij}\nabla_i\nabla_j K({\bf X},{\bf X}'; t) - W^T_{ij}X'_i\nabla_j K({\bf X},{\bf X}'; t) -\frac{1}{2}{X}'_i\cdot U_{ij} {X}_j'K({\bf X},{\bf X}'; t),\label{fks}
\end{equation}
which corresponds to the stochastic differential equation
\begin{equation}
dX_i = \sqrt{D}_{ij}dB_j + W_{ij}X_j dt
\end{equation}
and the potential $V({\bf X}') = \frac{1}{2}{X}'_i\cdot U_{ij} {X}_j'$. Note that this agrees with the form for the corresponding Green's function written down in \cite{freed} for the semi-flexible polymer actions with higher order derivatives.
Making an {\sl Ansatz} of the form
\begin{equation}
K({\bf X},{\bf X}'; t)= Q(t)\exp(-\frac{1}{2} {\bf X}\cdot A(t){\bf X} -\frac{1}{2} {\bf X}'\cdot A'(t){\bf X}' - {\bf X}\cdot B(t){\bf X}').
\end{equation}
we find  that the above form is indeed a solution, the Gaussian ansatz can be motivated from the Markov property of the path integral and its quadratic nature, see for instance ref \cite{dean2007}. However the resulting equations for
$Q(t)$ and the coefficient matrices $A(t)$, $A'(t)$ and $B(t)$ are nonlinear and coupled, so no explicit form for these coefficients is apparent and we do not even write the appropriate equations down. We do note, however, that this strategy can be pushed through for simple path integrals without the second order derivative term, even in the case of certain time dependent potentials \cite{pop1969,grit2010,dean2009,dean2010,boy2011,boy2012,dean2019}.
We also note then when ${\bf X}={\bf X}'={\bf 0}$ the prefactor $Q$ is related to a functional determinant, and some general results exist in the mathematics literature regarding this term \cite{duke}.
While the general Gaussian solution is thus difficult to write down, we next look for a solution of the Feynman-Kac equation  at late times, or in the {\em ground state dominance} case if one considers the quantum mechanics context. We look for solutions  of the form
\begin{equation}
K({\bf 0},{\bf X}'; t) = d'_R\exp(-\mu_R t-\frac{1}{2} {\bf X}'\cdot S_R{\bf X}'),
\label{ansatz}
\end{equation}
where the subscript $R$ signifies that the argument is the coordinate on the right of the
propagator and $d'_R$ is an undetermined normalization constant. The {\sl Ansatz} of Eq. (\ref{ansatz}) has the physical interpretation of the partition function for a bulk system of large length  $t$ with the variables ${\bf X}$ fixed at a boundary with a vacuum. The term $\mu_R$ would correspond to the ground state energy in the context of quantum mechanics or to the bulk free energy per unit length in the context of statistical mechanics.
Inserting the {\sl Ansatz} Eq. (\ref{ansatz}) into Eq. (\ref{fks})   we obtain
\begin{equation}
\mu_R = \frac{1}{2}{\rm Tr}DS_R
\end{equation}
along with
\begin{equation}
{\bf X}'\cdot(\frac{1}{2}S_RDS_R +W^TS_R -\frac{1}{2}U){\bf X'}=0.
\end{equation}
The latter of these two equations  thus yields the {\sl algebraic Riccati equation}
\begin{equation}
S_RDS_R +W^TS_R+ S_RW -U=0.\label{ricr}
\end{equation}
The backward equation has the solution
\begin{equation}
K({\bf X},{\bf 0}; t) = d'_L\exp(-\mu_L t-\frac{1}{2} {\bf X}\cdot S_L{\bf X}),
\end{equation}
where the subscript $L$ now indicates that it is the left argument of the propagator that is varying. The resulting equation for $S_L$ is then
\begin{equation}
S_LDS_L -W^TS_L- S_LW -U=0, \label{ricl}
\end{equation}
and we find
\begin{equation}
\mu_L = \frac{1}{2}{\rm Tr}DS^T_L.
\end{equation}

In the particular case of interest here we
have
\begin{equation}
D_{ij} = \delta_{ij}\delta_{i,n-1},
\qquad {\rm and} \qquad
W_{ij} = \delta_{i+1,j}.\label{defW}
\end{equation}

Intuitively, from a time reversal argument, we should find that
\begin{equation}
K({\bf X}, {\bf X}';t) = K(P{\bf X}', P{\bf X};t),
\end{equation}
where the {\sl parity operator} $P_{ij} = (-1)^i\delta_{ij}$ changes the sign of all the terms $d^ix(s)/ds^i$ having
an odd number of time derivatives. To see this from the algebraic Riccati equations we act
on Eq. (\ref{ricl}) on the left and right by $P$ and use the fact that $P^2 =I$ to write
\begin{equation}
PS_LPPDPPS_LP -PW^TPPS_LP- PS_LPPWP -PUP=0.
\end{equation}
Now we note that for an arbitrary matrix $A$
\begin{equation}
[PAP]_{ij}= (-1)^{i+j}A_{ij}.
\end{equation}
From this we see that $PDP=D$ and $PUP=U$ as $D$ and $U$ are both diagonal, while $PWP=-W$ and  $PW^TP=-W^T$. This then gives
\begin{equation}
PS_LPDPS_LP +W^TPS_LP+ PS_LPW -U=0,
\end{equation}
which is the same equation as Eq. (\ref{ricr}) and so we find
\begin{equation}
PS_LP=S_R.
\end{equation}
Using this we also find that $\mu_R=\mu_L=\mu$, which should physically be the case given the thermodynamic interpretation of $\mu_L$ and $\mu_R$ in terms of a free energy per unit length.

\section{Unconfined path integrals}\label{gaus}
The key point of the method applied here is the evaluation of the {\em unconfined} path integral
\begin{equation}
K_u({\bf X},{\bf X}';t) = \int d[x] \delta({\bf X}(0)-{\bf X})\delta({\bf X}(t)-{\bf X}')
\exp\left(-\frac{1}{2}\int_{-\infty}^\infty  ds \sum_{k=0}^n a_k \left[\frac{d^kx(s)}{ds^k}\right]^2\right).
\end{equation}
As in the case of the propagators we wish to study here, the fields ${\bf X}(s)$ are fixed at the times $0$ and $t$, but crucially the energy term in the exponential is present at all times and not just in the interval $[0,t]$. This means that the propagators $K_u$ and $K$ are mathematically different and correspond to different physical systems. To use statistical physics terminology, in the confined case there is no bulk or field outside the region $[0,t]$ while in the unconfined case there is.

We first assume that as $s\to\pm\infty$ the fields ${\bf X}(s)\to {\bf 0}$ and we can thus integrate by parts to obtain
\begin{equation}
K_u({\bf X},{\bf X}';t) = \int d[x] \delta({\bf X}(0)-{\bf X})\delta({\bf X}(t)-{\bf X}')
\exp\left(-\frac{1}{2}\int_{-\infty}^\infty  ds \sum_{k=0}^n  x(s)\left[a_k(-1)^{k}\frac{d^{2k}}{ds^{2k}}\right]x(s)\right).
\end{equation}
Written in this way we see that, taken with the appropriate normalization, $K_u({\bf X},{\bf X}';t)$ is simply the joint probability distribution for
the Gaussian random variables $(x(0), \ \dot x(0), \cdots x^{(n-1)}(0))$ (around the point ${\bf X}$) and $(x(t), \ \dot x(t), \cdots x^{(n-1)}(t))$ (around the point ${\bf X}'$) where the field $x(s)$ has correlation function
\begin{equation}
\langle x(s)x(s')\rangle =G(s-s')
\end{equation}
and $G$ is the Green's function satisfying
\begin{equation}
\sum_{k=0}^n \left[a_k(-1)^{k}\frac{d^{2k}}{ds^{2k}}\right]G(s-s')=\delta(s-s').
\end{equation}
Defining the correlation matrix $C$ by
\begin{equation}
C= \left\langle \begin{pmatrix} & {\bf X}(0) \\ &{\bf X}(t)\end{pmatrix} \begin{pmatrix} & {\bf X}(0) \\ &{\bf X}(t)\end{pmatrix}^T\right\rangle
\end{equation}
we thus find
\begin{equation}
K_u({\bf X},{\bf X}';t) = \frac{1}{(2\pi)^n [{\rm det}(C)]^{\frac{1}{2}}}\exp\left(-\frac{1}{2} \begin{pmatrix} & {\bf X} \\ &{\bf X}'\end{pmatrix}\cdot C^{-1} \begin{pmatrix} & {\bf X} \\ &{\bf X}'\end{pmatrix}\right),
\end{equation}
which is the normalized joint probability density function for Gaussian random variables.
Note that the correlation matrix can be written in  block form as
\begin{equation}
C = \begin{pmatrix}& C_0 & C_t\\ & C^T_t & C_0\end{pmatrix},
\end{equation}
where
\begin{equation}
C_t = \langle {\bf X}(0){\bf X}^T(t)\rangle.
\end{equation}
Using standard results on block matrices we may write
\begin{eqnarray}
C^{-1} &=& \begin{pmatrix}& (C_0 -C_t C_0^{-1}C^T_t)^{-1} & -C_0^{-1}C_t (C_0 -C^T_t C_0^{-1}C_t)^{-1}\\ & -(C_0 -C^T_t C_0^{-1}C_t)^{-1}C_t^TC_0^{-1} & (C_0 -C^T_t C_0^{-1}C_t)^{-1}\end{pmatrix} \quad {\rm with} \quad
{\rm det} (C) = {\rm det}( C_0^2 - C^T_t C_0^{-1}C_t C_0),\nonumber\\
~
\end{eqnarray}
yielding
\begin{eqnarray}
&&K_u({\bf X},{\bf X}';t) = \frac{1}{(2\pi)^n [{\rm det}(C_0^2 - C^T_t C_0^{-1}C_t C_0)]^{\frac{1}{2}}}\times\nonumber \\&&
\exp(-\frac{1}{2}{\bf X}\cdot(C_0- C_t C_0^{-1}C^T_t)^{-1} {\bf X}-\frac{1}{2}{\bf X}'\cdot(C_0- C^T_t C_0^{-1}C_t)^{-1} {\bf X}'
+ {\bf X}\cdot\left[ C_0^{-1} C_t (C_0- C^T_t C_0^{-1}C_t)^{-1}\right]{\bf X}')).
\end{eqnarray}
If we consider the initial condition $t=0$ of the unrestricted propagator, we see that
\begin{equation}
K_u({\bf X},{\bf X}';0) =\delta({\bf X}-{\bf X}')p({\bf X})\label{icku},
\end{equation}
where $p({\bf X})$ is the probability density function of ${\bf X}(t)$ at any point $t$. This is thus simply given by
\begin{equation}
p({\bf X}) = \frac{1}{(2\pi)^{\frac{n}{2}}[{\rm det}(C_0)]^{\frac{1}{2}}}
\exp(-\frac{1}{2} {\bf X}\cdot C_0^{-1} {\bf X}).
\end{equation}
The elements of the correlation matrices are computed using the invariance under translation of the Greens function
\begin{equation}
[C_t]_{n'm'} = \langle \frac{d^{n'} x(t')}{dt'^{n'}}\frac{d^{m'} x(t)}{dt^{m'}}|_{t'=0}\rangle = \frac{d^{n'} }{dt'^{n'}}\frac{d^{m'} }{dt^{m'}}G(t-t')|_{t'=0} = (-1)^{n'} \frac{d^{n'+m'} }{dt^{n'+m'}}G(t),
\end{equation}
where the indices run from $0$ to $n-1$.
From this one can verify that
\begin{equation}
C_t^T= PC_tP.
\end{equation}
This result, along with $P^2=I$,  allows us to write
\begin{equation}
(C_0- C_t C_0^{-1}C^T_t)^{-1} = P(C_0- C^T_t C_0^{-1}C_t)^{-1}P,
\end{equation}
and thus
\begin{eqnarray}
&&K_u({\bf X},{\bf X}';t) = \frac{1}{(2\pi)^n [{\rm det}(C_0^2 - C^T_t C_0^{-1}C_t C_0)]^{\frac{1}{2}}}\times\nonumber \\&&
\exp(-\frac{1}{2}{\bf X}\cdot P(C_0- C^T_t C_0^{-1}C_t)^{-1}P {\bf X}-\frac{1}{2}{\bf X}'\cdot(C_0- C^T_t C_0^{-1}C_t)^{-1} {\bf X}'
+ {\bf X}\cdot\left[ C_0^{-1} C_t (C_0- C^T_t C_0^{-1}C_t)^{-1}\right]{\bf X}')).
\end{eqnarray}
The Green's function can be computed in a straightforward manner, choosing
{$a_n=1$} we can write
\begin{equation}
 \sum_{k=0}^n  a_k(-1)^{k}\frac{d^{2k}}{ds^{2k}}=
 \prod_{k=1}^n \left[-\frac{d^2}{ds^2} + \omega_k^2\right],
\end{equation}
so we write the operator in terms of its fundamental frequencies $\omega_k$. Fourier transforming then gives
\begin{equation}
\tilde G(\omega) =\frac{1}{\prod_{k=1}^n (\omega^2 + \omega_k^2)}\label{gff}
\end{equation}
and upon inverting the Fourier transform we obtain
\begin{equation}
G(t) = \sum_{k=1}^n \frac{\exp(-\omega_k |t|)}{2\omega_k \prod_{1\leq i\neq k\leq n} (\omega_i^2 -\omega_k^2)}.\label{gf}
\end{equation}
In the above we choose the roots $\omega_k$ to have a positive real part.
From these results the unconfined propagator can be derived in terms and
the matrix inversion required can be easily carried out using computer algebra.

An alternative way of analysing the correlation matrix $C$ is to consider the stochastic differential equation leading to the same correlation function $G(t)$. The stochastic process can  simply be generated by the coupled set of equations
\begin{eqnarray}
{dX_0} &=& X_1 dt\nonumber\\
{dX_1} &=& X_2 dt \nonumber\\
\cdots &=&\cdots\nonumber \\
{dX_{n-1}}&=& - \sum_{k=0}^{n-1} b_k X_k dt +dB_t,
\end{eqnarray}
where $X_k = d^k x/dt^k$. The above process is obviously Gaussian and its correlation function in Fourier space is
\begin{equation}
\tilde G'(\omega) = \frac{1}{ \sum_{k=0}^{n} b_k (i\omega)^k \sum_{k=0}^{n} b_k (-i\omega)^k},
\end{equation}
if we define $b_n=1$. Now if we take
\begin{equation}
\sum_{k=0}^{n} b_k (i\omega)^k = \prod_{k=1}^n (i\omega + \omega_k),
\end{equation}
we recover the correlation function $\tilde G$ of Eq. (\ref{gff}). We note in particular that $b_{n-1} = \sum_{k=1}^n \omega_k$.
The stochastic equation for this process may be written as
\begin{equation}
dX_i = -H_{ij}X_jdt + \sqrt{D_{ij}}dB_j,
\end{equation}
where $D_{ij}$ is as defined in  Eq. (\ref{defW}) and
\begin{equation}
H_{ij} = -W_{ij} +\delta_{n-1i}b_j.
\end{equation}

Another way of analysing the properties of the correlation matrix $C$, rather than using the
Green's function representation above, is to integrate the corresponding Langevin equation
\begin{equation}
\frac{d{\bf X}}{dt} = -H{\bf X} + \boldsymbol{\eta}(t),
\end{equation}
where $\boldsymbol{\eta}$ is Gaussian white noise with correlation function
\begin{equation}
\langle \boldsymbol{\eta}  (t)\boldsymbol{\eta}^T  (t')\rangle = \delta(t-t')D.
\end{equation}
Starting from ${\bf X}={\bf 0}$ at $t=0$ we find the solution

\begin{equation}
{\bf X}(t) = \int_{0}^t ds \exp\left(-(t-s) H\right) \boldsymbol{\eta}(s)
\end{equation}

Using this we compute the equal time correlation function { $C(t,t) = \langle
{\bf X}(t){\bf X}(t)^T\rangle$}, finding

\begin{equation}
C(t,t)= \int_{0}^t ds\exp\left(-(t-s) H\right)D\exp\left(-(t-s) H^{T}\right).
\end{equation}
Taking the time derivative of the above gives
\begin{equation}
\frac{d C(t,t)}{dt} = -HC - CH^T + D.
\end{equation}
Defining $C_0=\lim_{t\to\infty} C(t,t)$ which is independent of time, we then obtain
\begin{equation}
HC_0 +C_0H^T = D.\label{lya}
\end{equation}

\section{Determination of the propagator from the unconfined propagator}\label{trans}
In order to compute the confined propagator we formally use it to compute the unconfined propagator that we found in the previous section. It is clear that we can write
\begin{equation}
K_u({\bf X},{\bf X'};t) = \lim_{L\to\infty}K({\bf 0},{\bf X}; L)K({\bf X},{\bf X}';t)K({\bf X}',0; L-t).
\end{equation}
The two additional propagators on the right hand side above insert the energy of the field into the regions $(-\infty,0]$  and $[t,\infty)$ to give the
unconfined propagator where the field energy is present at all times. In the preceding section, we have seen that the Feynman-Kac formula gives
\begin{equation}
K({\bf 0},{\bf X}; L) = d'_R \exp(-\mu L -\frac{1}{2}{\bf X}\cdot S_R {\bf X})
= d_R \exp( -\frac{1}{2}{\bf X}\cdot S_R {\bf X}),
\end{equation}
and
\begin{equation}
K({\bf X},{\bf 0};L- t) = d'_L\exp(-\mu (L-t) -\frac{1}{2}{\bf X}\cdot S_L {\bf X})
= d_L\exp(\mu t -\frac{1}{2}{\bf X}\cdot S_L {\bf X})
\end{equation}
where the divergent  $\exp(-\mu L)$ is absorbed into the constants $d_L$ and $d_R$.
This then yields
\begin{equation}
K({\bf X},{\bf X'};t) = \frac{1}{d_R d_L}\exp(-\mu t)K_u({\bf X},{\bf X'};t)\exp( \frac{1}{2}{\bf X}\cdot S_R {\bf X})\exp( \frac{1}{2}{\bf X}'\cdot S_L {\bf X}').
\end{equation}
Using the initial conditions for $K_u({\bf X},{\bf X'};t)$ in Eq. (\ref{icku}) and $K({\bf X},{\bf X'};t)$ gives
\begin{equation}
1 = \frac{1}{d_R d_L}\frac{1}{(2\pi)^{\frac{n}{2}}[{\rm det}(C_0)]^{\frac{1}{2}}}
\exp(-\frac{1}{2} {\bf X}\cdot C_0^{-1} {\bf X})\exp( \frac{1}{2}{\bf X}\cdot S_R{\bf X}+ \frac{1}{2}{\bf X}\cdot S_L {\bf X}).
\end{equation}
This gives the relation
\begin{equation}
{\bf X}\cdot C_0^{-1} {\bf X}= {\bf X}\cdot[S_R+ S_L ]{\bf X}
\end{equation}
and so, as all the matrices above are symmetric,
\begin{equation}
S_R+ S_L = C_0^{-1}.
\end{equation}
From this we also find that
\begin{equation}
\mu = \frac{1}{4}[{\rm Tr} DS_R^T + {\rm Tr} DS_L^T] = \frac{1}{4}{\rm Tr} DC_0^{-1}=
\frac{1}{4}[C_0^{-1}]_{n-1\ n-1}.\label{muc}
\end{equation}
The above relation thus gives some partial information on the solution of the algebraic Riccati equations Eqs. (\ref{ricl}) and (\ref{ricr}) but unfortunately not the solution of all the matrix elements.

Note that from  Eq. (\ref{lya}) we can write
\begin{equation}
H+  C_0H^TC_0^{-1} = DC_0^{-1}
\end{equation}
and taking the trace we find from Eq. (\ref{muc})
\begin{equation}
\mu = \frac{1}{4}{\rm Tr} DC_0^{-1} = \frac{1}{2}{\rm Tr} H = \frac{1}{2}b_{n-1} =\frac{1}{2} \sum_{k=1}^n \omega_k.\label{genmu}
\end{equation}
This shows that the bulk free energy per unit length is just the sum of the contributions from each fundamental frequency. We also find that
\begin{equation}
d_R d_L = \frac{1}{(2\pi)^{\frac{n}{2}}[{\rm det}(C_0)]^{\frac{1}{2}}}.
\end{equation}
Putting all this together we obtain
\begin{eqnarray}
&&K({\bf X},{\bf X'};t) = (2\pi)^{\frac{n}{2}}[{\rm det}(C_0)]^{\frac{1}{2}}\exp(-\mu t)\exp( \frac{1}{2}{\bf X}\cdot S_R {\bf X})\exp( \frac{1}{2}{\bf X}'\cdot S_L {\bf X}')\times\frac{1}{(2\pi)^n [{\rm det}(C_0^2 - C^T_t C_0^{-1}C_t C_0)]^{\frac{1}{2}}}\times\nonumber \\&&
\exp(-\frac{1}{2}{\bf X}\cdot P(C_0- C^T_t C_0^{-1}C_t)^{-1} P{\bf X}-\frac{1}{2}{\bf X}'\cdot(C_0- C^T_t C_0^{-1}C_t)^{-1} {\bf X}' + {\bf X}\cdot\left[ C_0^{-1} C_t (C_0- C^T_t C_0^{-1}C_t)^{-1}\right]{\bf X}'),
\end{eqnarray}
which simplifies to give
\begin{eqnarray}
&&K({\bf X},{\bf X'};t) = \frac{\exp(-\mu t)}{(2\pi)^{\frac{n}{2}} [{\rm det}(C_0 - C^T_t C_0^{-1}C_t )]^{\frac{1}{2}}}
\exp(-\frac{1}{2}{\bf X}\cdot P[(C_0- C^T_t C_0^{-1}C_t)^{-1}-S_L]P {\bf X}\nonumber \\&-&\frac{1}{2}{\bf X}'\cdot[(C_0- C^T_t C_0^{-1}C_t)^{-1} -S_L]{\bf X}'
+ {\bf X}\cdot\left[ C_0^{-1} C_t (C_0- C^T_t C_0^{-1}C_t)^{-1}\right]{\bf X}')).\label{master}
\end{eqnarray}
The above expression thus gives the full solution for the propagator $K({\bf X},{\bf X}',t)$ in terms of the correlation matrix $C(t)$ and the solution of the Riccati equation (\ref{ricl}).

\section{Kleinert's second order path integral}\label{klein}
Here we rederive the second order path integral of Kleinert \cite{klein86,handbook} by the methodology described above. In the notation introduced
earlier we have
\begin{equation}
{\bf X} = \begin{pmatrix} &x\\ &v\end{pmatrix}
\end{equation}
\begin{equation}
U = \begin{pmatrix}& \omega_1^2\omega_2^2 & 0 \\
&0 & \omega_1^2+\omega_2^2\end{pmatrix}, \ \ W = \begin{pmatrix} &0 &1\\ & 0 & 0\end{pmatrix},  \ \ D = \begin{pmatrix} &0 &0\\ & 0 & 1\end{pmatrix}.
\end{equation}
The solution of the algebraic Riccati equation (\ref{ricr}) gives
\begin{equation}
S_R =\begin{pmatrix} &\omega_1\omega_2(\omega_1+\omega_2) & -\omega_1\omega_2 \\
& -\omega_1\omega_2  & \omega_1 +\omega_2\end{pmatrix}, \ S_L=PS_RP= \begin{pmatrix} &\omega_1\omega_2(\omega_1+\omega_2) & \omega_1\omega_2 \\
& \omega_1\omega_2  & \omega_1 +\omega_2\end{pmatrix}
\end{equation}
where we have taken the roots $\omega_1$ and $\omega_2$ to have positive real part
in order to ensure that the diagonal terms of $S_R$ are  positive. There are in fact
four solutions in total, which can be obtained by changing the signs of $\omega_1$ and $\omega_2$ in the formula above.  We also find that
\begin{equation}
\mu = \frac{1}{2}(\omega_1 + \omega_2),
\end{equation}
in agreement with the general formula of Eq. (\ref{genmu}).
This result is particularly interesting and merits some comment.
We can apply the method developed here to compute the propagator for the
simple Harmonic oscillator (first order) path integral. In this case the Feynman-Kac equation is simply the Euclidean Schr\"odinger equation and
its Gaussian solution is the ground state wave function of the simple harmonic oscillator. The bulk free energy per unit length is simply $\frac{1}{2} \omega$, the corresponding ground state energy. We see that for the second order path integral, the bulk free energy per unit length is the sum of the ground state energies of the oscillators corresponding to the two fundamental frequencies $\omega_1$ and $\omega_2$.
Computing the correlation matrix $C_t$ in Eq. (\ref{master}) we find
\begin{equation}
C_0^{-1} = 2\begin{pmatrix}&  \omega_1 \omega_2 (\omega_1 + \omega_2), &0\\
&0 & \omega_1 + \omega_2\end{pmatrix},
\end{equation}
and so we see the relation $C_0^{-1} = S_R + S_L$ is explicitly verified. We also find
\begin{equation}
{\rm det}(C_0 - C^T_t C_0^{-1}C_t ) = \exp(-t(\omega_1+\omega_2))\frac{M}{\omega_1 \omega_2 (\omega_1^2 -\omega_2^2)^2},
\end{equation}
where
\begin{equation}
M = (\omega_1^2+\omega_2^2) s_1 s_2 - 2\omega_1 \omega_2 c_1 c_2+2\omega_1\omega_2,
\end{equation}
and we have used the notation $s_i = \sinh(\omega_i t)$ and $c_i =\cosh(\omega_i t)$.
We thus obtain
\begin{eqnarray}
K(x,v,x',v';t) &=& \frac{(\omega_1\omega_2)^\frac{1}{2}[(\omega_1^2-\omega_2^2)^2]^\frac{1}{2}}{2\pi\left((\omega_1^2 + \omega_2^2)s_1 s_2 -2 \omega_1\omega_2 c_1 c_2 + 2\omega_1 \omega_2)\right)^\frac{1}{2}}\times \nonumber \\&&
\exp\left(-\frac{1}{2}\begin{pmatrix}&x'\\&v'\end{pmatrix}\cdot S_D \begin{pmatrix}&x'\\&v'\end{pmatrix} -\frac{1}{2}\begin{pmatrix}&x\\&-v\end{pmatrix}\cdot S_D \begin{pmatrix}
&x\\&-v\end{pmatrix} + \begin{pmatrix}&x\\&v\end{pmatrix}\cdot S_C \begin{pmatrix}&x'\\&v'\end{pmatrix}\right),\label{kleinert}
\end{eqnarray}
where
\begin{equation}
S_D= \frac{1}{M}\left(
\begin{array}{cc}
 \omega_1\omega_2 (\omega_1^2-\omega_2^2) (\omega_1 c_2 s_1-\omega_2 c_1 s_2) &  \omega_1\omega_2
   \left( 2\omega_1 \omega_2s_1 s_2- \left(\omega_1^2+\omega_2^2\right) (c_1c_2-1)\right) \\ \omega_1\omega_2
   \left( 2\omega_1 \omega_2s_1 s_2- \left(\omega_1^2+\omega_2^2\right) (c_1c_2-1)\right)
   & (\omega_1^2-\omega_2^2)(\omega_1 c_1s_2-\omega_2 c_2  s_1) \\
\end{array}
\right)
\end{equation}
and
\begin{equation}
S_C= \frac{\omega_1^2 -\omega_2^2}{M}\begin{pmatrix}&\omega_1\omega_2(\omega_1 s_1 -\omega_2 s_2) & -\omega_1\omega_2(c_1-c_2) \\ & \omega_1\omega_2(c_1-c_2) & \omega_1 s_2 -\omega_2 s_1\end{pmatrix}.
\end{equation}
This reproduces exactly the result of Kleinert \cite{klein86,handbook} apart from the constant factor $[(\omega_1^2-\omega_2^2)^2]^\frac{1}{2}$ which is given in \cite{klein86,handbook} as $|\omega_1^2-\omega_2^2|$. In the case where $\omega_1$ and $\omega_2$ are real the two versions are equivalent. However when $\omega_1$ and $\omega_2$ are imaginary and conjugate,  one must use the version given here.

\section{Extension to third order path integral}\label{higher}
Here we examine how the methodology described above can be extended to third order path integrals. The terms given in our general formula Eq. (\ref{master}) can all be computed by standard matrix algebra with the exception of the algebraic Riccati equation
Eq. (\ref{ricr}). The solution to this equation is to determine the relation between the confined and unconfined path integral
problem and in what follows we examine the forms of $S_R$ and $S_L$ for third order path integrals. The remaining terms in the formula can be expressed in terms of the correlation matrix but the resulting expressions are too long and unwieldy to present here and the formal expressions given are of use for computer aided algebra.

Here the relevant variables are the position $x$, velocity $v$ and acceleration $a$ encoded in the vector
\begin{equation}
{\bf X} = \begin{pmatrix}&x\\ &v \\&a \end{pmatrix}.
\end{equation}

For this third order theory the solution of the Riccati equation with positive terms on the diagonal is
\begin{equation}
S_R = \begin{pmatrix}
&\omega_1 \omega_2 \omega_3 (\omega_1 \omega_2 + \omega_1 \omega_3 + \omega_2 \omega_3)&-\omega_1 \omega_2 \omega_3 (\omega_1 + \omega_2 + \omega_3)&\omega_1 \omega_2 \omega_3\\
&-\omega_1 \omega_2 \omega_3 (\omega_1 + \omega_2 + \omega_3) &(\omega_2 + \omega_3) (\omega_1+\omega_3)(\omega_1+\omega_2) &-\omega_1 \omega_2 - \omega_1 \omega_3 - \omega_2 \omega_3\\
&\omega_1 \omega_2 \omega_3 & -\omega_1 \omega_2 - \omega_1 \omega_3 - \omega_2 \omega_3& \omega_1 + \omega_2 + \omega_3
\end{pmatrix}.
\end{equation}
From this we explicitly see that the bulk free energy $\mu$ is given by
\begin{equation}
\mu = \frac{1}{2}(\omega_1 + \omega_2 + \omega_3),
\end{equation}
so in accordance with our general results,  the free energy per unit length is given by the sum of the individual oscillator free energies.
The matrix $S_L$ is given by
\begin{equation}
S_L = \begin{pmatrix}
&\omega_1 \omega_2 \omega_3 (\omega_1 \omega_2 + \omega_1 \omega_3 + \omega_2 \omega_3)&\omega_1 \omega_2 \omega_3 (\omega_1 + \omega_2 + \omega_3)&\omega_1 \omega_2 \omega_3\\
&\omega_1 \omega_2 \omega_3 (\omega_1 + \omega_2 + \omega_3) &(\omega_2 + \omega_3) (\omega_1+\omega_3)(\omega_1+\omega_2) &\omega_1 \omega_2 +\omega_1 \omega_3 + \omega_2 \omega_3\\
&\omega_1 \omega_2 \omega_3 & \omega_1 \omega_2  +\omega_1 \omega_3  +\omega_2 \omega_3& \omega_1 + \omega_2 + \omega_3
\end{pmatrix}.
\end{equation}
Computing the inverse correlation matrix gives
\begin{equation}
C^{-1}_0 = \begin{pmatrix}& 2 \omega_1 \omega_2 \omega_3 (\omega_2 \omega_3 + \omega_1 \omega_2 + \omega_1\omega_3)&0&2 \omega_1 \omega_2 \omega_3\\
&0&2 (\omega_1 + \omega_2) (\omega_1 + \omega_3) (\omega_2 + \omega_3)&0\\
&2 \omega_1 \omega_2 \omega_3&0& 2 (\omega_1 + \omega_2 + \omega_3)
\end{pmatrix},
\end{equation}
so that again the equality $S_L+S_R = C_0^{-1}$ is satisfied.

\section{Conclusions}
We have show how the propagators of higher order path integrals can be calculated by exploiting a link with the two point functions of Gaussian random fields over all time. The key point was to represent the unconfined  Gaussian problem via the confined path integral. Due to the presence of the field throughout space, the Gaussian random field problem has an underlying spatial translational invariance which considerably simplifies its analysis.
The propagators are then related by
\begin{itemize}
\item[(i)] multiplication of a {\em bulk like} ground state wave function which is Gaussian in nature and the relevant matrix is shown to obey
an algebraic Riccati equation, and
\item[(ii)] a bulk like energy term $\exp(-\mu t)$ where
$\mu = \frac{1}{2}\sum_{k=1}^n \omega_k$ where the $\omega_k$ are the fundamental frequencies of the problem.
\end{itemize}
This bulk term is absent in the unconfined problem as it corresponds to a bulk pressure which is cancelled by the presence of the field outside the interval $[0,t]$.

The results derived above are relevant in several contexts, where it is known that the physical description leads to higher order derivative field actions that can be reduced to the calculation of the associated one-dimensional functional integrals. In particular, it should be helpful in the context of the (pseudo)Casimir interactions for higher order field Lagrangians, as is the case for the, {\sl e.g.}, Brazovskii media \cite{uchida}, where it can deliver explicit results for the separation dependence of the thermal fluctuation driven interaction for confined fields bounded by two semi-infinite non-fluctuating media, or for unconfined fields in the case of two infinitely thin surfaces immersed in the space pervading bulk.

From a mathematical point of view we have seen that there are some interesting links with the control theory due to the appearance of algebraic Riccati equations. During our study we have also found in Eq. (\ref{genmu}) an intriguing result about the statistics of
derivatives of Gaussian random field and the consequent form of the bulk energy of these
higher derivative one dimensional field theories. There are many different methods of evaluating first order path integrals, in particular there is a method based on the relationship between quadratic functionals of Brownian motion and Brownian local time via the  Ray-Knight theorem \cite{chan1994,dean1995}.  Brownian local time corresponds physically to monomer density, it would be interesting to see if such a relationship extends to higher
order path integrals.

\section{Acknowledgments}
D.S.D. acknowledges support from the ANR Grants FISICS and RaMaTraF.
R.P. would like to acknowledge the support of the 1000-Talents Program of the Chinese Foreign Experts Bureau, and by the University of the Chinese Academy of Sciences, Beijing.  B.M. acknowledges support from the National Natural
Science Foundation of China (NSFC) (Grant Nos. 21774131 and 21544007).

\end{document}